\documentstyle[12pt,epsfig]{article}
\begin{document}
\centerline{\Large\bf Reply to:``Comment to: `Corrections to the fine
structure} 
\centerline{\Large\bf constant in the spacetime of a cosmic string from}
\centerline{\Large\bf the generalized uncertainty principle' ''}
\vspace*{0.025truein}
\centerline{Forough Nasseri\footnote{Email: nasseri@fastmail.fm}}
\centerline{\it Physics Department, Khayyam Planetarium,
P.O.Box 769, Neishabour, Iran}
\begin{center}
(\today)
\end{center}

\begin{abstract}
In this Reply, using G. de A. Marques' comment, we correct calculations
and results presented in [Phys. Lett. B 632 (2006) 151-154]
about corrections to the fine structure constant in the spacetime
of a cosmic string from the generalized uncertainty principle.
\end{abstract}

In a recent Letter \cite{1}, we calculated the Bohr radius in the
spacetime of a cosmic string. Our calculations to obtain the Bohr
radius in the spacetime of a cosmic string were based on the
quantization of the angular momentum in units of $\hbar$ i.e.
$L_n=n \hbar$.

Based on G. de A. Marques' comment in Ref.\cite{2},
the quantization of $L_n=n \hbar$ is correct only for a flat spacetime.
In other words, in the presence of a cosmic string the quantization
of the angular momentum must be considered as $L_{n_{(b)}}=\frac{n}{b} \hbar$
where $b= 1 - \frac{4 G \mu}{c^2}$ (see Ref. \cite{2}).
As pointed out in Eq.(7) of Ref.\cite{2},
in the weak field approximation we have
\begin{equation}
\label{1}
\frac{a_B}{{\hat a}_B} \approx 1 - \frac{G \mu}{c^2}
\left( 8 + \frac{\pi}{4} \right),
\end{equation}
where $a_B$ is the Bohr radius in the absence of a cosmic string
and ${\hat a}_B$ is the Bohr radius in the presence of a cosmic string.

The purpose of this Reply is to correct calculations of Ref.\cite{1}
in accordance with the comment presented in Ref.\cite{2}.
Inserting Eq.(\ref{1}) in the effective Planck constant
${\hat \hbar}_{\rm eff}$ in the
spacetime of a cosmic string,
as given in Eq.(30) of Ref.\cite{1},
one can easily obtain 
\begin{equation}
\label{2}
{\hat \hbar}_{\rm eff} \simeq \hbar \Bigg[ 1 + {\hat \beta}^2
\left( \frac{m e^2
\left[ 1 -  \frac{G \mu}{c^2}( 8 + \frac{\pi}{4} )  \right] }
{4 \pi \epsilon_0 M_P^3 G } \right)^2 \Bigg],
\end{equation}
where we use
\begin{equation}
\label{p}
\frac{L_P}{{\hat a}_B}=
\frac{me^2 \left[ 1- \frac{G \mu}{c^2} \left( 8 + \frac{\pi}{4} \right) \right]}
{4 \pi \epsilon_0 M_P^3 G} \ll 1.
\end{equation}
Eq.(\ref{p}) is the corrected form of Eq.(31) in Ref.\cite{1}.
Substituting Eq.(\ref{2}) of this Reply in Eq.(34) of Ref.\cite{1},
we are led to
\begin{eqnarray}
\label{3}
{\hat \alpha}_{\rm eff} & \simeq & \left( \frac{e^2}{4 \pi \epsilon_0 \hbar c}
-\frac{\pi}{4} \frac{G \mu}{c^2} \frac{e^2}{4 \pi \epsilon_0 \hbar c} \right) \nonumber\\
& \times & \Bigg[ 1
- {\hat \beta}^2
\left( \frac{m e^2
\left[ 1 -  \frac{G \mu}{c^2}( 8 + \frac{\pi}{4} )  \right] }
{4 \pi \epsilon_0 M_P^3 G } \right)^2 \Bigg].
\end{eqnarray}
This equation is the corrected form of Eq.(35) in Ref.\cite{1}.
One can rewrite Eq.(\ref{3}) as
\begin{eqnarray}
\label{4}
{\hat \alpha}_{\rm eff} & \simeq & \left( \frac{e^2}{4 \pi \epsilon_0 \hbar c}
-\frac{\pi}{4} \frac{G \mu}{c^2} \frac{e^2}{4 \pi \epsilon_0 \hbar c} \right) \nonumber\\
& \times & \left\{ 1
- {\hat \beta}^2
\times 9.30 \times 10^{-50} \left[ 1- 2 \times
\left( 8 + \frac{\pi}{4} \right)
\times 10^{-6} \right] \right\},
\end{eqnarray}
where we use $\left[ 1- \left( 8 + \frac{\pi}{4} \right)
\frac{G \mu}{c^2} \right]^2 \simeq
\left[ 1-2 \times \left( 8 + \frac{\pi}{4} \right)
\frac{G \mu}{c^2} \right]$ and
$\frac{G \mu}{c^2} \sim 10^{-6}$.

It must be emphasized that Eq.(\ref{4}) of this Reply is the
corrected form of Eq.(36) in Ref.\cite{1}.
As given in Eq.(33) of Ref.\cite{1}, the fine structure constant
${\hat \alpha}$ in the spacetime of a cosmic string is defined by
${\hat \alpha} := \alpha \left( 1 - 
\frac{\pi G \mu}{4 c^2} \right)$.
Using Eq.(33) of Ref.\cite{1} and Eq.(\ref{4}) of this Reply,
we obtain
\begin{equation}
\label{5}
{\hat \alpha}_{\rm eff} \simeq {\hat \alpha} \left\{ 1
- {\hat \beta}^2 \times 9.30 \times 10^{-50} \left[ 1- 2 \times
\left( 8 + \frac{\pi}{4} \right) \times
10^{-6} \right] \right\}.
\end{equation}
Indeed, Eq.(\ref{5}) is the corrected form of Eq.(37) of Ref.\cite{1}.
The only differences between Eq.(\ref{4}) of this Reply and
Eq.(36) of Ref.\cite{1}, and also between Eq.(\ref{5}) of this Reply
and Eq.(37) of Ref.\cite{1},
are due to the number $8$ in the expression
$\left( 8 + \frac{\pi}{4} \right)$
as appears in  Eqs.(\ref{4}) and (\ref{5}).
In the weak field approximation,
these differences have arisen from the quantization
$L_{n_{(b)}}=\frac{n}{b} \hbar$ where $b=1 - \frac{4 G \mu}{c^2}$
as pointed out in Ref.\cite{2} and as considered in the calculations of
this Reply.

Finally, it must be emphasized on the importance of this Reply
because the author of the Comment \cite{2} has obtained only the
effects of the quantization of $L_{n_{(b)}}=\frac{n}{b} \hbar$ on the
Bohr radius in the spacetime of a cosmic string and has not
calculated the effects of this quantization 
on the main results of Ref.\cite{1} which are
the corrections to the fine structure constant in the
spacetime of a cosmic string from the generalized uncertainty principle.
Our calculations in this Reply complete the calculations presented
in the Comment \cite{2}.\\
{\bf Acknowledgements:} F.N. thanks Hurieh Husseinian and A.A. Nasseri
for noble helps and also thanks Amir and Shahrokh for truthful helps.
It is a pleasure to thank N. Pileroudi for sending me the pdf file
of Ref.\cite{2}.

\end{document}